\documentclass[11pt]{article}
\usepackage[T1]{fontenc}
\usepackage{amsmath,amssymb,amsthm}
\usepackage[margin=1in]{geometry}
\usepackage{booktabs}
\usepackage{enumitem}
\usepackage[hidelinks]{hyperref}
\usepackage{xurl}
\newtheorem{theorem}{Theorem}

\newtheorem{lemma}{Lemma}

\newtheorem{corollary}[theorem]{Corollary}
\theoremstyle{definition}

\theoremstyle{remark}
\newtheorem{remark}[lemma]{Remark}
\newcommand{\C}{\mathbb{C}}
\newcommand{\Q}{\mathbb{Q}}
\newcommand{\Z}{\mathbb{Z}}
\newcommand{\F}{\mathbb{F}}
\newcommand{\PP}{\mathbb{P}}
\newcommand{\Fp}{\overline{\mathbb{F}}_p}

\title{No complete set of mutually unbiased bases in dimension ten\\
contains the Fourier pair or three product bases}
\author{Jinze Yang$^{1}$, Yile Liu$^{2}$, Yangyang Li$^{3,*}$ and Xiu-Hao Deng$^{4,5,\dagger}$\\[1.5ex]
\small\itshape $^{1}$School of Physics, Xidian University, Xi'an 710071, China\\
\small\itshape $^{2}$School of Mathematical Sciences, Peking University, Beijing 100871, China\\
\small\itshape $^{3}$Key Laboratory of Intelligent Perception and Image Understanding of Ministry of Education of China,\\
\small\itshape Collaborative Innovation Center of Quantum Information of Shaanxi Province,\\
\small\itshape Institute of Interdisciplinary Quantum Science and Technology (IIQST),\\
\small\itshape School of Artificial Intelligence, Xidian University, Xi'an 710071, China\\
\small\itshape $^{4}$Shenzhen International Quantum Academy, Shenzhen 518048, China\\
\small\itshape $^{5}$Shenzhen Branch, Hefei National Laboratory, Shenzhen 518048, China\\[1ex]
\small $^{*}$\href{mailto:yyli@xidian.edu.cn}{yyli@xidian.edu.cn}\qquad
$^{\dagger}$\href{mailto:dengxiuhao@iqasz.cn}{dengxiuhao@iqasz.cn}}
\date{\today}

\begin{document}
\maketitle

\begin{abstract}
With computer-assisted proofs that $\C^6$ admits at most three mutually unbiased bases (MUBs) now available, $\C^{10}$ is the smallest
dimension in which it is unknown whether a complete set of $d+1=11$ MUBs exists; only three MUBs are known. We prove
that every triple of mutually unbiased product bases of $\C^2\otimes\C^5$ is strongly unextendible: no vector is
unbiased to all three bases. Consequently no set of four MUBs in dimension ten, and in particular no complete set,
contains three product bases. This settles the case $d=2\times5$ anticipated by McNulty, Pammer and Weigert; it was
known only for the Heisenberg--Weyl triple. The proof covers all ten equivalence classes with explicit Nullstellensatz
certificates modulo one prime on $91$ charts of $\PP^9\times\PP^9$ for each class ($910$ in all), transferred to characteristic zero by a
projective reduction lemma, together with an independent rank certificate for eight of the ten classes. We also prove that the pair formed by the standard and the Fourier basis of $\C^{10}$ is contained in no set of four
MUBs. Exactly $3040$ rays are unbiased to both bases (this confirms the count found numerically by F\"uhr and Rzeszotnik), each of them is unbiased to at most eight of the others, and every triple formed by the pair and a third
MUB is strongly unextendible; these statements were announced by Grassl and Sz\"oll\H{o}si in 2022, and we give a
complete proof. For this we certify $34{,}940$ cyclic $10$-roots in Krawczyk uniqueness boxes, decide
unimodularity exactly through an involution, and prove completeness by a four-dimensional interval branch-and-bound
obtained from the Chinese-remainder splitting $\Z_{10}=\Z_2\times\Z_5$.
\end{abstract}

\section{Introduction}

Two orthonormal bases of $\C^d$ are \emph{mutually unbiased} if every inner product between a vector of one and a
vector of the other has modulus $d^{-1/2}$. At most $d+1$ bases of $\C^d$ can be pairwise mutually unbiased, and
complete sets of $d+1$ mutually unbiased bases (MUBs) exist whenever $d$ is a prime power
\cite{Ivanovic1981,WoottersFields1989}. Let $N(d)$ denote the maximal size of a set of MUBs in $\C^d$. For every other
$d$ the value of $N(d)$ was open, and the smallest case, $d=6$, resisted for decades \cite{DEBZ2010}. In the autumn of
2026 several computer-assisted proofs of $N(6)=3$ appeared
\cite{OAIMUB,Jaswal2026a,CWT2026,Jaswal2026b}. Dimension ten is now the smallest dimension in which $N(d)$ is unknown:
$3\le N(10)\le 11$. By Weiner's gap theorem \cite{Weiner2013}, $N(10)\le 9$ is equivalent to the non-existence of
eleven MUBs.

Two families of configurations have shaped the work on dimension six, and both have natural analogues in dimension ten.
\begin{itemize}[leftmargin=*]
\item \emph{Product bases.} For $\C^6=\C^2\otimes\C^3$, McNulty and Weigert classified all mutually unbiased product
  bases, showed that no triple of them extends, and that a complete set contains at most one product basis
  \cite{McNultyWeigert2012a,McNultyWeigert2012b,McNultyWeigert2012c}.
  For $\C^2\otimes\C^5$, McNulty, Pammer and Weigert showed that there are at most three mutually unbiased product
  bases and brought every such triple into a normal form \cite[Thm.~2, Cor.~3]{MPW2016}. Recalling that in dimension six
  no vector is unbiased to three mutually unbiased product bases, they wrote: ``We expect similar results to hold for
  larger product dimensions such as $d=2\times5$, but we have not been able to generalize the proof for $d=6$''
  \cite[after Cor.~6]{MPW2016}; see also \cite{WPZ2011,Cariello2021}. For $\C^{10}$, Grassl, McNulty, Mi\v{s}ta and Paterek showed
  by computer algebra that the Heisenberg--Weyl triple $\{Z,X,XZ\}$ is strongly unextendible \cite{GMMP2017}
  (see also \cite{MBGW2014}). The recent review by McNulty and Weigert asks whether the sets of mutually unbiased product bases
  giving the lower bound on $N(d)$ in composite dimensions are unextendible, and records only $d=6$ as solved
  \cite[Problem~10.8]{MW2026}.
\item \emph{The Fourier pair.} Grassl found that exactly $48$ vectors are unbiased to both the standard basis $I$ and the
  Fourier basis $F_6$, that they form $16$ bases, and that no further vector is unbiased to $I$, $F_6$ and any of them
  \cite{Grassl2004}, \cite[Thm.~7.6]{MW2026}; hence no four MUBs contain $(I,F_6)$ (see also \cite{BBELTZ2007}, who
  note that Bj\"orck and coworkers had found the same vectors). Jaming, Matolcsi, M\'ora, Sz\"oll\H{o}si and Weiner
  extended this to the two-parameter Fourier family \cite{JMMSW2009}. The vectors unbiased
  to $I$ and $F_d$ are the biunimodular sequences of length $d$, equivalently the unimodular cyclic $d$-roots
  \cite{Bjorck1990,BjorckFroberg1991,Haagerup2008,FuhrRzeszotnik2015}. For $d=10$ the cyclic system has $34{,}940$ roots, a count
  obtained by Gröbner-basis computations modulo primes and by numerical homotopy methods
  \cite{Faugere2002,FaugereSvartz2013,SteelMagma,SVW2001}. F\"uhr and Rzeszotnik found numerically $3040$ unimodular
  ones in ten symmetry orbits \cite{FuhrRzeszotnik2015}, and Magsino and Xu found the same number \cite{MagsinoXu2023}.
  In a 2022 talk, Grassl and Sz\"oll\H{o}si reported, from modular Gr\"obner-basis computations, $3040$ unimodular
  cyclic $10$-roots, $1136$ complex Hadamard matrices obtained from them, and that all resulting triples with $I$ and
  $F_{10}$ are strongly unextendible \cite{GrasslSzollosi2022}; no proof has appeared. Sun recently studied
  completions of Fourier-type pairs in dimension $2m$ \cite{Sun2026}.
\end{itemize}

We settle both questions in dimension ten.
A set of bases is \emph{strongly unextendible} if no unit vector is unbiased
to all of its bases. We identify $\C^{10}=\C^2\otimes\C^5$; a \emph{product basis} is an orthonormal basis consisting
of product vectors $a\otimes b$.

\begin{theorem}[product triples]\label{thm:A}
Every triple of pairwise mutually unbiased product bases of $\C^2\otimes\C^5$ is strongly unextendible.
\end{theorem}

In particular, Theorem~\ref{thm:A} answers \cite[Problem~10.8]{MW2026} for $d=10$, in the strong sense.

\begin{theorem}[the Fourier pair]\label{thm:B}
Let $I$ be the standard basis and $F_{10}$ the Fourier basis of $\C^{10}$. Then:
\begin{enumerate}[label=(\roman*)]
\item exactly $3040$ rays are unbiased to both $I$ and $F_{10}$;
\item each of these rays is unbiased to at most $8$ of the others; at least $2240$ of them are unbiased to none;
\item there are no four MUBs in $\C^{10}$ containing $I$ and $F_{10}$;
\item there are at most $1136$ bases unbiased to both $I$ and $F_{10}$, and for every such basis $B$ the triple
  $\{I,F_{10},B\}$ is strongly unextendible.
\end{enumerate}
\end{theorem}

Theorem~\ref{thm:B}(iv) is the ten-dimensional analogue of Grassl's result for $(I,F_6)$. Theorem~\ref{thm:B}(i)
and (iv) prove the count $3040$ and the strong unextendibility announced by Grassl and Sz\"oll\H{o}si
\cite{GrasslSzollosi2022}, and our bound of $1136$ third bases matches their $1136$ Hadamard matrices; the method uses
neither the number of cyclic $10$-roots nor modular computation.

\begin{corollary}\label{cor:C}
No set of four MUBs in $\C^{10}$, and in particular no complete set, contains
(a) three bases that are product bases for one tensor factorisation $\C^{10}\cong\C^2\otimes\C^5$, or
(b) a pair of bases equivalent to $(I,F_{10})$.
\end{corollary}

Here two pairs of bases are equivalent if one common unitary or antiunitary map carries the first pair to the second,
up to phases, the order of the vectors and the order of the two bases; such maps preserve all moduli of inner
products. Corollary~\ref{cor:C}(a) follows from Theorem~\ref{thm:A} because a fourth MUB would consist of vectors
unbiased to all three product bases, and (b) is Theorem~\ref{thm:B}(iii). Together with \cite[Thm.~2]{MPW2016}, (a)
lowers the upper bound on the number of product bases (for one factorisation) in a set of four or more MUBs in
$\C^2\otimes\C^5$ from three to two. Theorem~\ref{thm:A} is the statement expected in
\cite[after Cor.~6]{MPW2016}. Damski and \.Zyczkowski recently constructed $p+1$ MUBs in dimension $pq$ for primes
$p<q$, the standard basis and $p$ maximally entangled bases, and left open whether these sets extend
\cite{DamskiZyczkowski2026}. For $(p,q)=(2,5)$ their three bases are the image of a triple of mutually unbiased
product bases under a common permutation of coordinates, so the triple is strongly unextendible by
Theorem~\ref{thm:A}; it also contains a pair equivalent to $(I,F_{10})$, to which Theorem~\ref{thm:B} applies.

The two results meet in one configuration. Under the Chinese-remainder identification $\Z_{10}\cong\Z_2\times\Z_5$ the
pair $(I,F_{10})$ is a pair of product bases, and the $40$ product rays unbiased to it are the Gaussian chirps
$e^{\pi i(cn^2+2bn)/10}/\sqrt{10}$ with $c\in\{1,3,7,9\}$, $b\in\Z_{10}$ (Remark~\ref{rem:chirps}). Theorem~\ref{thm:A} excludes a
fourth basis for every triple of product bases. Theorem~\ref{thm:B} shows that for this particular
pair \emph{every} third basis, product or not, already blocks a fourth.

Neither result decides whether $N(10)\le 9$: a complete set need contain neither the Fourier pair nor three product
bases. We return to this point in Section~\ref{sec:discussion}.

\paragraph{Certification standard.} Every computational step that the proofs use is exact (integer arithmetic, in
particular modulo one prime, with identities that an independent program re-expands) or rigorous interval arithmetic
with outward rounding. Non-rigorous floating-point computations were used only to find certificates and starting
points, never to validate them. The root enclosures, the unimodularity decision and the overlap graphs were computed
twice, with independent code and arithmetic, and the graphs agree bit for bit. Two steps have a single implementation:
the clique enumeration on the conservative graphs (exact combinatorics) and the completeness step of
Theorem~\ref{thm:B} (Section~\ref{sec:complete}). The logic of an earlier version of the latter was audited
independently, and the audit's blocking findings were fixed; a second independent audit then asked for two further
safeguards, which the final version has: interval operations return the whole real line when an operand is not
finite, so that a floating-point overflow can never cause a false exclusion, and the whole proof is carried out by
one build, so that a single catalogue of root boxes is used (Section~\ref{sec:complete}). The program splits into
$256$ independent subproblems and records each completion. Theorem~\ref{thm:A} uses as external input only the
classification of \cite{MPW2016}, with the normalisation step from \cite{BWB2010}; Theorem~\ref{thm:B} uses none.

\section{\texorpdfstring{Product triples in $\C^2\otimes\C^5$}{Product triples in C2 x C5}}\label{sec:A}

\subsection{Classification}
Let $\zeta=e^{2\pi i/20}$, $\omega=\zeta^4$ and $i=\zeta^5$. The bases of $\C^5$ that occur are $G_\infty=I_5$ and
$G_t$, $t\in\F_5$, with entries $G_t(x,j)=\omega^{tx^2+jx}/\sqrt5$; together they form the complete set of six MUBs of
$\C^5$. The qubit vectors are the eigenvectors $q_{b,0},q_{b,1}$ of the Pauli operators $b\in\{Z,X,Y\}$. A
\emph{label assignment} $\ell=(\ell_{Z,0},\ell_{Z,1},\ell_{X,0},\ell_{X,1},\ell_{Y,0},\ell_{Y,1})$ with labels in
$\PP^1(\F_5)=\{\infty,0,1,2,3,4\}$ defines three product bases
\begin{equation}\label{eq:U}
U_b=\bigl[\,q_{b,0}\otimes G_{\ell_{b,0}},\ q_{b,1}\otimes G_{\ell_{b,1}}\bigr],\qquad b\in\{Z,X,Y\}.
\end{equation}
They are pairwise mutually unbiased if and only if the three label sets $\{\ell_{b,0},\ell_{b,1}\}$ are pairwise
disjoint; the two labels within one pair may coincide.

By the classification of McNulty, Pammer and Weigert \cite[Cor.~3]{MPW2016}, every triple of mutually unbiased product
bases of $\C^2\otimes\C^5$ is equivalent to one of the form \eqref{eq:U}. Their equivalences include local complex
conjugations. On the normal forms these are realised by global maps: since $\overline{G_t(:,j)}=G_{-t}(:,-j)$ and
$M_2G_t(:,j)=G_{-t}(:,3j)$ with $M_2|x\rangle=|2x\rangle$, conjugation of the $\C^5$ factor alone acts on the three
bases as the unitary $I_2\otimes M_2$, and conjugation of the qubit factor alone acts as the antiunitary
$(I_2\otimes M_2)\mathcal K$, where $\mathcal K$ is global complex conjugation. A simultaneous normalisation step uses the
fact that in $\C^5$ every pair of MUBs has a unique completion and the only rays unbiased to $I_5$ and $F_5$ are the
$20$ columns of $G_1,\dots,G_4$ \cite[Sec.~4.1]{BWB2010}. Hence strong unextendibility, which is invariant under common
unitary or antiunitary maps, phases and permutations of vectors, needs to be checked only on the assignments
\eqref{eq:U}.

There are $\sum_{r=0}^3\binom3r(6)_{6-r}=4080$ admissible assignments, where $r$ is the number of bases whose two
labels coincide and $(n)_k=n(n-1)\cdots(n-k+1)$. The qubit Clifford group (generated by $H$ and
$S$, acting on the six qubit rays), the action of $\mathrm{PSL}(2,5)$ on the labels (generated by $F_5$, $D=\mathrm{diag}
(\omega^{x^2})$ and $M_2$), and global conjugation act on them by explicit integer permutations. The $4080$ assignments
form ten orbits (Table~\ref{tab:orbits}); we checked this by enumerating all $6^6$ six-tuples with integer arithmetic.

\begin{table}[ht]
\centering
\begin{tabular}{rllllr}
\toprule
orbit & $Z$ labels & $X$ labels & $Y$ labels & size & second proof\\
\midrule
0 & $\infty,\infty$ & $0,0$ & $1,1$ & 60 & --- \\
1 & $\infty,\infty$ & $0,0$ & $2,2$ & 60 & --- \\
2 & $\infty,\infty$ & $0,0$ & $1,2$ & 720 & rank 975\\
3 & $\infty,\infty$ & $0,0$ & $1,4$ & 180 & rank 975\\
4 & $\infty,\infty$ & $0,0$ & $2,3$ & 180 & rank 975\\
5 & $\infty,\infty$ & $0,1$ & $2,3$ & 720 & rank 975\\
6 & $\infty,\infty$ & $0,1$ & $2,4$ & 720 & rank 975\\
7 & $\infty,\infty$ & $0,2$ & $1,3$ & 720 & rank 975\\
8 & $\infty,0$ & $1,2$ & $3,4$ & 480 & rank 987\\
9 & $\infty,0$ & $1,4$ & $2,3$ & 240 & rank 987\\
\bottomrule
\end{tabular}
\caption{The ten classes of mutually unbiased product triples in $\C^2\otimes\C^5$. Orbit~$0$ is the Heisenberg--Weyl
triple of \cite{GMMP2017}. All ten are covered by the chart certificates of Section~\ref{sec:charts}; the last column
refers to the independent rank certificate of Section~\ref{sec:rank}.}
\label{tab:orbits}
\end{table}

\subsection{Exact coefficients}
Write $T_{t,s}(j,k)=\sqrt5\,(G_t^*G_s)_{jk}$ for $t\neq s$ and let $\chi$ be the quadratic character of $\F_5$.
Completing the square in the quadratic Gauss sum gives
\begin{equation}\label{eq:gauss}
\begin{gathered}
T_{t,\infty}(j,k)=\omega^{-tk^2-jk},\qquad T_{\infty,s}(j,k)=\omega^{sj^2+kj},\\
T_{t,s}(j,k)=\chi(s-t)\,\omega^{-(k-j)^2/(4(s-t))}\qquad(s\neq t\in\F_5).
\end{gathered}
\end{equation}
For an assignment put $C=\sqrt{10}\,\bigl[\begin{smallmatrix}U_X^*U_Z\\ U_Y^*U_Z\end{smallmatrix}\bigr]$, a
$20\times10$ matrix. By \eqref{eq:gauss} each entry of $C$ is a twentieth root of unity $\zeta^{E_{ak}}$ with an
explicit integer exponent table $E$. All computations below start from these integer tables; no floating-point
quantity enters any certificate.

\subsection{Polynomial formulation}
Let $\psi$ be a unit vector unbiased to $U_Z,U_X,U_Y$ and put $z=\sqrt{10}\,U_Z^*\psi$. Unbiasedness to $U_Z$ says
$|z_k|^2=1$ for all $k$; unbiasedness to the $a$-th vector of $U_X$ or $U_Y$ says $|\sum_kC_{ak}z_k|^2=10$. Replacing
$\overline{z}$ by a second vector of unknowns $w$, the diagonal terms contribute $\sum_kz_kw_k=10$, so the conditions
become
\begin{equation}\label{eq:system}
z_kw_k=z_0w_0\quad(1\le k\le 9),\qquad
f_a(z,w)=\sum_{k\neq l}\zeta^{E_{ak}-E_{al}}\,z_kw_l=0\quad(1\le a\le 20).
\end{equation}
These equations are bihomogeneous of bidegree $(1,1)$ and define a closed subscheme $\mathcal V\subset\PP^9\times\PP^9$
over $\Z[\zeta,1/5]$. The point $([z],[\overline z])$ lies in $\mathcal V(\C)$. Hence Theorem~\ref{thm:A} follows once
$\mathcal V(\C)=\emptyset$ for the ten representatives.

\subsection{Reduction to one prime}
Fix $p=1073741741\equiv1\pmod{20}$ and the residue map $\zeta\mapsto a=95182621$, a primitive $20$th root of unity
modulo~$p$.

\begin{lemma}[projective reduction]\label{lem:A}
If $\mathcal V(\Fp)=\emptyset$ under $\zeta\mapsto a$, then $\mathcal V(\C)=\emptyset$.
\end{lemma}
\begin{proof}
A complex point gives a point over a number field $L\supseteq\Q(\zeta)$. Choose a prime $\mathfrak P$ of $L$ above the
prime $(p,\zeta-a)$. Scale each of the two homogeneous coordinate vectors so that all coordinates are
$\mathfrak P$-integral and one is a unit; the bihomogeneous equations are preserved. Reducing modulo $\mathfrak P$ gives
two nonzero vectors over a finite field of characteristic $p$ satisfying the reduced system.
\end{proof}

\begin{remark}[one prime suffices]
Reducing a system modulo a prime can in general lose solutions or create spurious ones, which is why modular
Gr\"obner-basis algorithms combine several primes or verify their output \cite{Arnold2003,MST2021}. We use only the
direction that is safe for projective schemes: a complex point reduces to a point over $\Fp$, so emptiness over $\Fp$
implies emptiness over $\C$. Solutions can disappear modulo $p$ only in affine settings, where coordinates may acquire
$p$ in their denominators; points modulo $p$ that do not lift play no role, since only emptiness is used. Moreover the
proof does not rely on a Gr\"obner computation being correct: each chart carries an explicit identity
$1=\sum_vh_vg_v$ over $\F_p$ that an independent program re-expands. An unlucky prime could therefore prevent a
certificate from existing, but could not produce a false one.
\end{remark}

\paragraph{Charts.} On $\mathcal V$, either $z_0w_0\ne0$, and we may scale to $z_0=w_0=1$ (the \emph{affine chart}:
$z_kw_k=1$ and the twenty $f_a$, in $18$ variables), or $z_0w_0=0$, in which case all $z_kw_k=0$; choosing
$z_i\neq0$ and $w_j\neq0$ forces $i\neq j$, and we scale to $z_i=w_j=1$, $w_i=z_j=0$ (the $90$ \emph{boundary charts}).
These $91$ charts cover $\mathcal V$. In each chart an identity $1=\sum_vh_vg_v$ over $\F_p$ among the chart
equations $g_v$ proves that the chart has no point over $\Fp$.

\subsection{The certificates}\label{sec:charts}
For each of the ten representatives and each of its $91$ charts we produced such an identity, $910$ in total.
\begin{itemize}[leftmargin=*]
\item The $900$ boundary identities were obtained with the \texttt{lift} command of \textsc{Singular} \cite{Singular}
  after \texttt{msolve} \cite{msolve} and \textsc{Singular} had both found the reduced Gröbner basis $\{1\}$.
\item For the ten affine charts \textsc{Singular}'s \texttt{lift} ran out of memory (at $6$\,GB per process;
  for orbit~$0$ also at $26$\,GB). We instead use the torus
  structure of the affine chart. Substituting $w_k=z_k^{-1}$ turns the generators into Laurent polynomials
  $\tilde f_a=\sum_{k\ne l}\zeta^{E_{ak}-E_{al}}z_k/z_l$ in $z_1,\dots,z_9$, and an identity
  $1=\sum_a c_a\tilde f_a$ with Laurent cofactors of bounded degree is found by solving a sparse Macaulay-type
  linear system modulo $p$, restricted to a symmetry-adapted subspace; this takes a few seconds and less than
  $0.1$\,GB per orbit (for orbits $0,1$ the system has $1789$ rows). Each Laurent identity is then rewritten exactly as a polynomial identity
  $1=\sum_kt_k(z_kw_k-1)+\sum_ah_af_a$ in the $18$ affine variables, of the same form as the \textsc{Singular}
  output.
\end{itemize}
The certificates were \emph{verified} independently of how they were found. A short program expands every product
$h_vg_v$ with integer arithmetic modulo $p$ and checks that the sum is the constant~$1$. It reads the chart equations
from inputs that a verification script regenerates from the exact exponent tables, and it requires that these inputs
coincide byte for byte with the inputs given to the certificate search. All $910$ identities pass (the largest affine
identity has $479{,}691$ cofactor terms). Modified certificates, such as swapped cofactors, a single altered
coefficient, or the certificate of one orbit applied to another orbit, are rejected.

\begin{proof}[Proof of Theorem~\ref{thm:A}]
By the classification, a unit vector unbiased to a mutually unbiased product triple gives, after a common unitary or
antiunitary map, a unit vector unbiased to one of the ten representatives, hence a complex point of the corresponding
$\mathcal V$. The $91$ verified chart identities show $\mathcal V(\Fp)=\emptyset$, and Lemma~\ref{lem:A} gives
$\mathcal V(\C)=\emptyset$, a contradiction.
\end{proof}

\subsection{An independent second proof for eight classes}\label{sec:rank}
Work in $R=\Q(\zeta)[c,z,w]/(z_kw_k-c)$, bigraded by $\deg z_k=(1,0)$, $\deg w_k=(0,1)$, $\deg c=(1,1)$. The Weyl
operators $I_2\otimes X_5$ and $I_2\otimes Z_5$ preserve each of $U_Z,U_X,U_Y$ as sets of rays and therefore permute the
$f_a$ up to phases; the resulting action of $W\cong\Z_5^2$ on the balanced components $R_{n,n}$ commutes with
reduction modulo~$p$, because the stabiliser orders $1,5,25$ are units. If the multiplication map
$\bigoplus R_{n-1,n-1}\to R_{n,n}^W$ by the $W$-projected generators has full row rank modulo $p$, a nonzero maximal
minor lifts to characteristic zero and $c^n\in\langle f_a\rangle$. Evaluating at $z=\sqrt{10}\,U_Z^*\psi$,
$w=\overline z$, $c=1$ gives $1=0$. At $n=3$ the matrices have full row rank $975$ (orbits $2$--$7$) and $987$
(orbits $8,9$); square witnesses are part of the data. For orbits $0,1$ the rank at $n=3$ is $972<975$, so this route
does not apply to them; they are covered by the chart certificates above.

\begin{remark}[Galois conjugate classes]
The chart equations depend on the exponent table only through the row differences $E_{ak}-E_{al}$. The automorphism
$\zeta\mapsto\zeta^{17}$ of $\Q(\zeta)$ maps the multiset of rows of the table of orbit~$0$, up to a phase per row, onto
that of orbit~$1$, and similarly orbit~$3$ onto orbit~$4$ (checked with exact integer arithmetic). The corresponding
systems are Galois conjugate, so their complex points correspond. Either member of each pair would therefore have
sufficed.
\end{remark}

\section{The Fourier pair}\label{sec:B}

Let $\omega_{10}=e^{2\pi i/10}$ and $F_{10}=(\omega_{10}^{jk}/\sqrt{10})_{j,k}$.

\subsection{\texorpdfstring{Cyclic $10$-roots}{Cyclic 10-roots}}
Let $\psi$ be unbiased to $I$ and $F_{10}$. Then $z=\sqrt{10}\,\psi$ has unimodular entries, and we fix the phase by
$z_0=1$. Unbiasedness to $F_{10}$ means $|\hat z_k|^2=\bigl|\sum_n z_n\omega_{10}^{-kn}\bigr|^2=10$ for all $k$.
Since $|\hat z_k|^2=\sum_s c_s\omega_{10}^{-ks}$ with the periodic autocorrelation $c_s=\sum_n z_{n+s}\overline{z_n}$
and $c_0=10$, this holds if and only if $c_s=0$ for $s=1,\dots,9$. Put $x_n=z_{n+1}/z_n$ (indices modulo~$10$). As
$\overline{z_n}=1/z_n$,
\[
c_s=\sum_{n}x_nx_{n+1}\cdots x_{n+s-1}=:C_s(x),\qquad \prod_n x_n=1 .
\]
Conversely, every root of the \emph{cyclic $10$-roots system}
\begin{equation}\label{eq:cyclic}
C_1(x)=\dots=C_9(x)=0,\qquad x_0x_1\cdots x_9=1
\end{equation}
with all $|x_n|=1$ defines $z_0=1$, $z_{n+1}=x_nz_n$, consistently around the cycle. This is Bj\"orck's
correspondence \cite{Bjorck1990}: rays unbiased to $I$ and $F_{10}$ are in bijection with unimodular roots of
\eqref{eq:cyclic}.

\subsection{The number of roots}\label{sec:count}
The cyclic $10$-roots system is known computationally to have $34{,}940$ roots: Gröbner-basis computations modulo the
prime $32003$ \cite{SteelMagma} and over a prime field (typically $\F_{65521}$) \cite{FaugereSvartz2013} give a zero-dimensional ideal with quotient of
dimension $34{,}940$, cyclic $10$-roots is a standard benchmark of the F5 algorithm \cite{Faugere2002}, numerical
irreducible decomposition finds $34{,}940$ isolated solutions \cite{SVW2001}, a Gröbner basis over $\Q$ has been
computed by multi-modular methods \cite{Parisse2019}, and Grassl and Sz\"oll\H{o}si announced a complete set of
algebraic solutions \cite{GrasslSzollosi2022}. We do not use this count: below we prove that there are at least
$34{,}940$ distinct roots, and in Section~\ref{sec:complete} we prove directly that the unimodular ones are exactly the
certified $3040$.

\subsection{\texorpdfstring{Certified enclosures of $34{,}940$ roots}{Certified enclosures of 34,940 roots}}\label{sec:enclosures}
We start from the $1747$ approximate representatives published with PHCpack \cite{PHCpack} and expand them by the
dihedral symmetry of \eqref{eq:cyclic} to $34{,}940$ approximate roots; these serve only as starting points. For each,
after a Newton polish, we apply Krawczyk's test \cite{Krawczyk1969,Moore1977,Rump1983}: with $\tilde x$ the approximate
root, $Y\approx J(\tilde x)^{-1}$ and a box $X\ni\tilde x$,
\[
K(X)=\tilde x-YF(\tilde x)+\bigl(I-YJ(X)\bigr)(X-\tilde x)\subset\operatorname{int}X
\]
implies that $X$ contains exactly one root, and that this root lies in $K(X)$. Complex rectangular interval
arithmetic on $\C^{10}$ encloses the real interval Jacobian on $\mathbb R^{20}$, so the real theorem applies. We ran
two independent implementations:
\begin{enumerate}[label=(\arabic*)]
\item a C++ program with double-precision intervals, in which every operation is rounded to nearest and then widened
  outward by one unit in the last place, compiled without floating-point contraction;
\item a SageMath program \cite{Sage} in which the system and its Jacobian are built symbolically over $\Q$ and
  evaluated in Arb ball arithmetic \cite{Arb} at $160$ bits, with dyadic boxes whose endpoints are exact.
\end{enumerate}
Both certify all $34{,}940$ boxes (the C++ enclosures have width below $6\times10^{-12}$), and both find the
enclosures pairwise disjoint. Hence \eqref{eq:cyclic} has at least $34{,}940$ distinct roots. The C++ program checks
at run time that its arithmetic is binary64 with gradual underflow and records the compiler; since every result is
widened outward by one unit in the last place, its enclosures remain valid under any IEEE rounding mode. Both programs
exit with an error unless all $34{,}940$ boxes
are certified, every root is classified, the enclosures are disjoint and the degree bound below holds.

\paragraph{Exact unimodularity.} The map $\sigma(x)=(1/\overline{x_n})_n$ permutes the roots of \eqref{eq:cyclic},
since $C_k(\sigma(x))=\overline{C_{10-k}(x)}\big/\overline{\textstyle\prod_nx_n}$ and $\prod_n\sigma(x)_n=1$. A root is
unimodular if and only if it is fixed by $\sigma$. If the interval image $\sigma(K)$ of an enclosure lies in the
uniqueness box $X$, the root is fixed, by uniqueness in $X$; if some enclosure of $|x_n|^2$ excludes $1$, it is not.
This decides every certified root: exactly $3040$ are unimodular and $31{,}900$ are not, in both implementations.

\subsection{All biunimodular vectors of length ten}\label{sec:complete}
We now show, without using the total number of roots, that the $3040$ certified rays are all rays unbiased to $I$ and $F_{10}$.

\paragraph{Reduction.} Identify $\Z_{10}=\Z_2\times\Z_5$ by $n\mapsto(n\bmod2,n\bmod5)$ and write $z_n=u_{n\bmod5}$ for
even $n$ and $z_n=v_{n\bmod5}$ for odd $n$. With $\omega_5=e^{2\pi i/5}$, $U_k=\sum_mu_m\omega_5^{-3km}$ and
$V_k=\sum_mv_m\omega_5^{-3km}$, the discrete Fourier transform of $z$ at the frequency $k\leftrightarrow(k_2,k_5)$ is
$U_{k_5}+(-1)^{k_2}V_{k_5}$. Hence $z$ is biunimodular if and only if
\begin{equation}\label{eq:crt}
|u_m|=|v_m|=1,\qquad \Re\bigl(U_k\overline{V_k}\bigr)=0,\qquad |U_k|^2+|V_k|^2=10\qquad(k\in\Z_5).
\end{equation}
If $U_k\neq0$ the last two conditions give $V_k=\varepsilon_k\,i\,U_k\,(10/|U_k|^2-1)^{1/2}$ with $\varepsilon_k=\pm1$
(boxes on which $U_k$ may vanish are treated separately in the search below).
After normalising $u_0=z_0=1$, a choice of $u_1,\dots,u_4$ on the unit circle and of signs $\varepsilon$ determines
$v$, and biunimodularity becomes the five equations $|v_m|^2=1$ in four real unknowns.

\paragraph{Symmetry.} A modulation $z_n\mapsto e^{2\pi ian/10}z_n$ fixes $z_0$ and maps $(U_k,V_k)$ to
$(U_{k-a},(-1)^aV_{k-a})$; the dilation $z_n\mapsto z_{3n}$ maps $U_k$ to $U_{2k}$. Since $\sum_k|U_k|^2=25$, every
biunimodular ray is the image of one with $|U_0|^2=\max_k|U_k|^2$, $|U_1|^2=\max_{k\ge1}|U_k|^2$ and $\varepsilon_0=+1$.
The certified set is closed under these maps; we check this by mapping the root enclosures and finding their images
inside certified uniqueness boxes.

\paragraph{Search.} Each $u_m$, $m=1,\dots,4$, is parametrised by two overlapping rational charts
$u=\pm\bigl((1-s^2)+2is\bigr)/(1+s^2)$, $s\in[-\tfrac{17}{16},\tfrac{17}{16}]$; with the $2^4$ sign patterns
$(\varepsilon_1,\dots,\varepsilon_4)$ this gives $16\cdot16=256$ four-dimensional subproblems. A box is excluded if an interval or mean-value enclosure (computed with interval automatic
differentiation) of some residual misses $0$, or if the symmetry inequalities fail on the whole box. Where the
direction of $U_k$ is not resolved, or $|V_k|$ is small, $V_k$ is computed instead from two of the linear equations
that $|v_m|^2=1$ imposes on it, and the conditions $|V_k|^2=10-|U_k|^2$ and $\Re(U_k\overline{V_k})=0$ are kept as
residuals. A box is covered if a Krawczyk test on a slightly inflated box certifies a unique zero of four of the
residuals, and that zero is either excluded by the remaining residuals or mapped by $x_n=z_{n+1}/z_n$ into a certified
uniqueness box of a cyclic $10$-root. Every other box is bisected. Every interval operation with a non-finite operand
returns the whole real line, so an overflow can only prevent an exclusion. All $256$ subproblems terminate with no
unresolved box. The certifying run was made in one piece on one workstation (one source file and one binary,
identified by their SHA-256 hashes; $16$ threads, $13.5$ hours): it processed $30{,}859{,}491{,}964$ boxes and
reported no overflowed enclosure. Its node count agrees, subproblem by subproblem, with earlier split runs of the
previous version on a server, a cluster and a workstation (about $2000$ core-hours with duplicates).

\paragraph{The catalogue.} The run certifies its own $3040$ uniqueness boxes from the input file (exactly unimodular,
with pairwise disjoint enclosures, closed under modulation and dilation), so all $256$ subproblems refer to one set of
$3040$ distinct biunimodular rays. The enclosures of Section~\ref{sec:enclosures} also give $3040$ distinct
unimodular roots; since every biunimodular ray lies in the first set, the two sets coincide.

\subsection{Overlap graphs}
For the $3040$ rays we enclose all overlaps $g(r,s)=|\langle\psi_r,\psi_s\rangle|^2$. Let $O$ and $U$ be the graphs of
genuinely orthogonal and genuinely unbiased pairs. Let $O^+$ join $r\neq s$ unless the enclosure of $g(r,s)$ excludes
$0$, and let $U^+$ join them unless it excludes $1/10$. Then $O\subseteq O^+$ and $U\subseteq U^+$. The enclosures have
width at most $1.1\times10^{-11}$, and the excluded pairs are far from the thresholds: every pair not in $O^+$ has
$g\ge8.0\times10^{-7}$, every pair not in $U^+$ has $|g-1/10|\ge6.1\times10^{-5}$, and no pair lies in both graphs.
$O^+$ has $57{,}680$ edges and $U^+$ has $3200$. The edge sets computed by the two implementations agree bit for bit.

A symmetry group of the pair, of order $1600$ \cite{FuhrRzeszotnik2015} (generated by the cyclic shift, the modulation $z_n\mapsto\omega_{10}^nz_n$, the
dilation $z_n\mapsto z_{3n}$, complex conjugation and the Fourier transform) acts on the rays; numerically the $3040$
rays form ten orbits of the sizes found in \cite{FuhrRzeszotnik2015}. Table~\ref{tab:fourier} lists the degrees in
the conservative graphs along this orbit decomposition. The proof uses only the degree bounds, not the orbits.

\begin{table}[ht]
\centering
\begin{tabular}{rrrl}
\toprule
orbit size & degree in $O^+$ & degree in $U^+$ & remark\\
\midrule
20, 20 & 144 & 0 & the $40$ product rays (Gaussian chirps)\\
200, 200 & 44 & 0 & \\
200 & 28 & 0 & \\
400, 400 & 52 & \textbf{8} & the only rays that may have unbiased partners\\
400, 400 & 28 & 0 & \\
800 & 28 & 0 & \\
\bottomrule
\end{tabular}
\caption{The $3040$ rays unbiased to $I$ and $F_{10}$, grouped by (numerically identified) symmetry orbit, with their
degrees in the conservative graphs $O^+\supseteq O$ and $U^+\supseteq U$. A degree $0$ in $U^+$ is exact; a positive
degree is an upper bound.}
\label{tab:fourier}
\end{table}

\begin{proof}[Proof of Theorem~\ref{thm:B}]
(i) follows from Section~\ref{sec:B}: the $3040$ certified unimodular roots are distinct, and by
Section~\ref{sec:complete} every biunimodular ray is one of them; (ii) is read off from $U^+\supseteq U$. For (iii), suppose $I,F_{10},B,D$ are MUBs. All
vectors of $B$ and $D$ are among the $3040$ rays, and for any $r\in B$ the ten rays of $D$ are unbiased to $r$, so
$r$ has degree at least $10$ in $U$, hence in $U^+$, contradicting (ii). For (iv), a basis $B$ unbiased to $I$ and
$F_{10}$ is a $10$-clique of the orthogonality graph $O\subseteq O^+$. An exhaustive enumeration finds
exactly $1136$ ten-cliques in $O^+$, so there are at most $1136$ such bases, and for each of these cliques the common
$U^+$-neighbourhood of its ten vertices is empty.
A vector unbiased to $I$, $F_{10}$ and $B$ would be one of the $3040$ rays and would lie in that common
neighbourhood.
\end{proof}

\begin{remark}\label{rem:chirps}
Under $n=5a+6x\pmod{10}$ the chirp $e^{\pi i(cn^2+2bn)/10}$ factorises as $(i^c(-1)^b)^a\,\omega^{4cx^2+3bx}$,
a product of a qubit vector unbiased to $Z$ and $X$ and a column of some $G_t$. These $40$ rays are exactly the
product rays unbiased to the pair; their mutual overlaps lie in $\{0,1/5,1\}$. Among the $1136$ candidate third bases, $16$ consist of product rays, and numerically the other $1120$ each lie inside
a single orbit. All $1136$ have pairwise inner products of modulus below $1.1\times10^{-15}$ numerically; we have not certified their exact
orthonormality, and Theorem~\ref{thm:B}(iv) does not need it. The candidate with ten rays from the eigenbasis of
$XZ$ recovers the Heisenberg--Weyl triple of \cite{GMMP2017}.
\end{remark}

\begin{remark}[second moments do not suffice]
A complete set is a complex projective $2$-design \cite{KlappeneckerRoetteler2005}, so the $90$ rays of the other nine
bases would satisfy $\sum_rw_r(|r\rangle\langle r|)^{\otimes2}=I+S-\sum_{v\in I\cup F_{10}}(|v\rangle\langle v|)^{\otimes2}$
with $w_r\in\{0,1\}$. The linear relaxation $0\le w_r\le1$ over the $3040$ rays is feasible numerically (residual $6\times10^{-11}$),
even with no weight on product rays. The obstruction of Theorem~\ref{thm:B} is therefore not visible at the level of the $2$-design identity;
it comes from the sparse unbiasedness graph.
\end{remark}

\section{Discussion}\label{sec:discussion}
Corollary~\ref{cor:C} removes two natural candidates for a complete set in $\C^{10}$, but it does not show
$N(10)\le9$. Since the set of pairs of bases that extend to four MUBs is closed (the set of quadruples of MUBs is
compact), Theorem~\ref{thm:B} also gives an open neighbourhood of $(I,F_{10})$ in which no pair extends to four MUBs;
making this neighbourhood explicit would be of interest. A complete set might contain neither the Fourier pair nor three product bases. Other pairs of the form
$(I,H)$ with $H$ a complex Hadamard matrix of order ten generally come in continuous families, whose unbiased vectors
need not be finite in number. In dimension six the result for $F_6$ \cite{Grassl2004} was later extended to the whole Fourier family
\cite{JMMSW2009} and eventually to all triples \cite{CWT2026,Jaswal2026b}. Extending Theorem~\ref{thm:B} from
$F_{10}$ to its affine families is the natural next step in dimension ten.

The certification techniques are standard (rigorous numerical certification was used for $d=6$ in \cite{JMMSW2009}
(discretisation with explicit error bounds) and \cite{CWT2026} (interval and ball arithmetic), and Gröbner-basis
arguments in \cite{Grassl2004,MBGW2014}); we do not claim novelty for
them. Two details may still be useful. Laurent Nullstellensatz certificates on the torus chart replace a
memory-bound Gröbner lift by a small sparse linear solve. Certified root enclosures, an involution that detects the unimodular locus exactly, and a low-dimensional interval
branch-and-bound from a Chinese-remainder splitting give finite, checkable lists of vectors unbiased to a pair whose
unbiased set is finite, without using the degree of the system.

\section*{Data availability}
The programs that generate the inputs and verify all certificates, the lists of SHA-256 checksums of every
certificate file, the logs of the verification runs, the logs of all branch-and-bound runs and the script that merges
them (it checks that every one of the $256$ subproblems finished without an unresolved box, under one source file and
one input file identified by their SHA-256 hashes), and the final completeness program with the complete log of its
certifying run are included as ancillary files of the arXiv
version. The programs and the certificates themselves ($910$ chart identities, eight rank witnesses and the root
enclosures; about $0.6$\,GB compressed) are deposited at Zenodo, \url{https://doi.org/10.5281/zenodo.24297026}.

\section*{Use of AI}
Claude Opus~5.5 was used for coding and for optimizing the speed of the algorithms.

\end{document}